\documentclass{article}
\usepackage{aip}
\usepackage{graphicx}
\usepackage{epsf}

\input{tcilatex}

\begin{document}

\title{On the nature of intermittency in weakly dissipative systems}
\author{F. Spineanu and M. Vlad \\
Association Euratom-Nasti (MER), \\
National Institute of Laser, Plasma and Radiation Physics\\
P.O. Box MG-36, Magurele, Bucharest, Romania}
\maketitle

\begin{abstract}
We propose a new perspective on the intermittency in weakly dissipative
systems. It is focused on the elementary (burst-like) event separating
states with different properties. This event is seen as a real-space-time
manifestation of an instanton connecting in an extended space-time states
with distinct topology of the idealized version of the system.
\end{abstract}

\section{Introduction}

The intermittency is in general associated with particular statistical
properties of a fluctuating quantity like the fields describing the
turbulent fluid or plasma. In certain regimes the nonlinear dynamics
generates cuasi-coherent structures like vortices and the high correlations
of these formations induce a non Gaussian statistics. The random generation
and destruction of cuasi-coherent vortices is one of the sources of
intermittency in fluid and plasma. The slow algebraic decay of the spectrum
at high wavenumbers is also considered a manifestation of the intermittency.
In this case the random events consisting of rapid energy exchanges at small
spatial scales should be attributed to vortex merging or island coalescence,
processes which generates quasi-singular layers at reconnection. For other
dynamical systems, the intermittency is manifested as a random sequence of
bursts, \emph{i.e.} time localised, strong increase of the amplitude of the
field. These events separate states where the system preserves stationary
properties, like are the regular flows or the turbulence with stationary
statistical properties. During such an event, which is very short in time,
the field is highly irregular, producing in general a reorganization of the
flow. To simplify the terminology we will say that the intermittent events
separate \emph{states with stationary properties} (SWSP) of the system.

We note that the statistical perspective on the intermittency is mainly
observational or attempts to characterize the statistical properties
(spectrum, scaling lows of the correlations) on the basis of the general
properties of the dynamical equations, like symmetry and conservation.

As an alternative to the statistical approach we propose in this paper a
general model for the intermittency, regarded as a dynamical process leading
to fast and significant changes in the fluctuating field. We take as the
essential characteristic of the intermittency the events consisting of
passages between states with stationary properties of the system. In many
cases it can be recognized, for ideal systems, a SWSP as a family of
configurations with the same topology. Consider for example a number of
vortices moving in an ideal fluid. Because the fluid is ideal it is not
possible to go from one topological configuration (for exemple $n$ vortices)
to a different one ($n^{\prime }\neq n$ vortices) since the passage requires
breaking and reconnection \ of the flow lines which in turn depends on the
presence of resistivity (in general, of dissipation). In real space-time
these transitions are not allowed and the ideal system is constrained to
evolve inside a family of topologically equivalent states, what we have
defined as a regime of SWSP. But even for the ideal system, if the
space-time is adequately extended it may be possible, under certain
conditions, that the transitions can be realized. The passage taking place
in an extended space-time between configurations with distinct topological
content will be called \emph{instanton}, in analogy with the instantons
connecting states of degenerate vacua in field theory.

Compared to the ideal case, the evolution of the \emph{weakly dissipative}
system consists of motions which are homotopically equivalent and, from time
to time, transitions between distinct configurations, the transitions being
only allowed due to the presence of the dissipation. The essential idea of
the model we propose is that these dissipative transitions are not
arbitrary: these transitions evolve in a way which is the manifestation in
real space-time of the instantons connecting topologically distinct
configurations in the extended space-time. These transitions in real
space-time are only possible if it exists for the particular system a
topological structure in which are embedded both the initial and the final
real space configurations.

\section{Intermittency and singularities in the plane of complex time}

Many numerical studies have been done for systems exhibiting intermittency
in the form of burst-like events. It has been found that there is a
connection between the positions of these events on the time axis and the
positions of the singularities of the solution in the complex time plane.
This has first been shown by Frisch and Morf \cite{FrischMorf} in a study of
a nonlinear differential equation for the variable (velocity) $v$ in the
presence of damping $\gamma $ and random drive $f\left( t\right) $: 
\begin{equation}
m\stackrel{.}{v}=-\gamma v-v^{3}+f\left( t\right)   \label{Frisch}
\end{equation}
The numerical results show that the singularity is located above the burst
event and that the amplitude is larger when the singularity is closer to the
real time axis. The connection has been investigated using the Fourier
representation of the solution.

This study presented a certain ressemblence with the problem of
integrability of second-order ordinary differential equation, which has been
formulated by Painlev\'{e} in the form of a precise criterion: a second
order differential equation is \emph{integrable} if the only movable
singularities of the solution, in the complex plane of the variable (\emph{%
e.g.} time), are poles. The analogy between the intermittency/singularity
and integrability/singularity problems has inspired a considerable effort of
characterizing by numerical methods the nature and the position in the
complex plane of the singularities, eventually leading to the extension of
the Painlev\'{e} criterion to larger classes of equations. Part of these
studies have been devoted to understanding chaos (via Melnikov integral) as
opposed to the integrability. More connected with the problem of
intermittency are the numerical studies of the following system \cite
{Bassetti}  \cite{Fucito}: 
\begin{equation}
\frac{\partial ^{2}\varphi }{\partial t^{2}}-\frac{\partial ^{2}\varphi }{%
\partial x^{2}}+m^{2}\varphi +g\varphi ^{3}=0  \label{bassetti}
\end{equation}
By numerical integration the position in the complex plane of the
singularities, denoted $x_{s}+iy_{s}$, has been obtained. In Bassetti \emph{%
et al. }\cite{Bassetti} the Fourier representation has been used to express
the function $\varphi $ by its singularities closest to the real time axis : 
$\func{Im}\varphi \sim \exp \left( -ky_{s}\right) $ and a graph of $%
y_{s}\left( t\right) $ is obtained numerically.

From the perspective explained above we have to see the two minima of the
potential in Eq.(\ref{bassetti}) as topologically distinct states \cite
{Rajaraman}. The transition between these minima is forbidden for a
classical particle, in the absence of negative dissipation (\emph{i.e.}
contact with a thermal bath). But in an extension of the space-time (here
simply : \emph{imaginary time}) there are \emph{instantons} (or topological
solitons) connecting the two states. In this particular case we can
precisely identify them, since the equation describes a $\varphi ^{4}$
theory and the solution is ($u$ is the velocity and $t_{0}$ is the initial
time) 
\[
\varphi \left( x,t\right) =i\frac{m}{\sqrt{g}}\tanh \left[ \frac{m}{\sqrt{2}}%
\frac{\left( t-t_{0}\right) -ux}{\sqrt{1-u^{2}}}\right] 
\]
This \emph{kink} is an instanton connecting the state $i\varphi =-m/\sqrt{g}$
at $t=-\infty $ with $i\varphi =+m/\sqrt{g}$ at $t=+\infty $. We take $u=0$
and $t_{0}=0$ and note that the $\tanh $ has singularities when the argument
is $\left( i\pi /2\right) \left( 2n+1\right) $, $n\in \mathbf{N}$. Using the
approximation $th\left( \frac{i\pi }{2}+x\right) \sim \frac{1}{x}$ we find
the imaginary part $y_{s}$ of the position in the complex $t$ plane of the
singularity: 
\[
y_{s}=-\frac{1}{k}\ln \left( \sqrt{\frac{g}{2}}t\right) 
\]
This formula reproduces the Fig.2 of Bassetti \emph{et al}.\cite{Bassetti}.
We have an example where the knowledge of the instanton explains the
connection between complex singularities and the intermittent bursts.

\section{The conjecture}

We formulate once more the basic idea \cite{FlorinMadi}  \cite{FlorinMadi1}:
the burst-like intermittency is a real-space/time manifestation of an
instanton transition between topologically distinct configurations
generalizing states with stationary properties (SWSP). The projection on
real-space/time is only allowed in the presence of dissipation.

This is nothing but a \emph{conjecture} and we have to gather solid
arguments to support it. However this idea already suggests a series of
steps to be taken in examining the model of an intermittent system.

\begin{enumerate}
\item  First, we have to find an ideal version of the system, \emph{e.g.} by
supressing the dissipation part. Actually this is often invoked in the
examination of the onset of chaos in systems weakly perturbed around
integrability.

\item  Then the \emph{ideal} model which is closest to the real one must be
examined for identifying the distinct topological classes to which the
solutions belong.

\item  Next we have to extend the system: not only the space-time must be
extended (to larger dimensionality and/or complex variables) but the
equation of the model must be embedded into a larger system (\emph{e.g.} the
simple pendulum equation is reduced from a self-dual Yang-Mills model).

\item  In this extended theory one should look for instantons connecting
configurations of distinct topological classes. The existence of these
solutions is the exclusive condition for the possibility of real space/time
intermittency.

\item  One should be able to find also the nature of singularities of these
instantons, whose signature could still be identifiable after returning to
the original real-space/time system. This will help to localize the bursts
by their relation with the extended space/time singularities.
\end{enumerate}

The extension of the theory and of the space-time means the inclusion of the
theory into a much larger context. For the ideal system it is necessary to
represent its SWSP as states with nontrivial topological content. The
instantons are transitions from one topological (SWSP) state to another.

In general one should expect that in arbitrary extension of the theory
instantons do not exist, due to topological constraints. The instanton must
be a solution of the extended system having as initial condition a
configuration with a particular topology and as target a configuration with
a different topology. Most frequently, the initial and final configurations
are sections of fibre bundles, as is the case for the $O\left( n\right) $
model. Connecting two configurations first requires to embed both homotopy
classes to which belong the two sections into one single object.

Of particular importance is to correctly infer the extended theory whose
equation must reduce to the original ones when we return to the real
space-time. There is however a series of deep connections that have been
revealed in recent years between classical integrable or topological
differential equations and the Self-Dual Yang-Mills (SDYM) field theory.
Reducing equations from SDYM has been done for many well known differential
equations, as will be discussed bellow. On the other hand there are precise
situations where the construction of instantons as geometric-algebraic
objects can be done systematically, using twistor theory. While these
instruments are very useful in investigating the conjecture, they do not
automatically lead to successful determination of the extended theory.

\section{Arguments supporting the conjecture}

The requirement to enlarge the theoretical framework of the \emph{ideal}
version of the system to a more complex theory can be formulated conversely:
the ideal system should result by reduction from a more general theory where
instantons can be found connecting the states of distinct topology. There is
a large number of ideal systems with exceptional integrability properties,
which can be derived by reduction from the Self-Dual Yang-Mills theory. For
eaxample the integrable hierarchies\cite{Mason1} (KdV, Nonlinear Schrodinger
Equation, \emph{sine}-Gordon, etc.), the Painlev\'{e} transcendents\cite
{Mason2}, etc. The self duality equations have solutions with nontrivial
topology inherited from the structure of fibre bundle defined by the base
space and the group of automorphism of the typical fibre, most frequently
(for principal fibre bundles) the algebra of symmetry group. Self duality
(equality of the curvature two form with its Hodge dual) provides nonzero
Chern class. It is in this framework that the first examples of instantons
(in particular the 't Hooft-Polyakov instanton) have been found. We shall
see bellow that one can construct instantons as Riemann surfaces, at least
for simple topology of the SWSP configurations.

\section{An example}

The simplest example of topological nontrivial state is represented by a
closed line on a torus surface. It is specified by two integer numbers $%
(m,n) $ which means that $m$ turns must be made in the toroidal direction
and respectively $n$ in the poloidal direction for the line to close in
itself. Any homotopic deformation preserves $\left( m,n\right) $. A line
with a different pair $\left( m^{\prime },n^{\prime }\right) $ cannot be
deformed into the first one, \emph{i.e.} the homotopy classes are labeled by
the two integers. There are homotopic deformations which makes that a line
from one family becomes close on a finite space region to a deformed line
belonging to other family. Then the presence of dissipation can allow for
reconnection, which produces a transition of the line from one family to
another. This occurs in real space-time. According to the \emph{conjecture},
this transition is the manifestation in real space-time of the existence of
a solution connecting in extended space-time solutions belonging to the two
families. We have to look for this \emph{super}solution.

To examine this example we simplify taking $m=1$. Any solution represents a
section in a fibred space whose base space is the circle (the axis of the
torus) and the space of internal symmetry (the fibre) is also a circle (a
phase variable, represented by an arrow from the current point on basis to a
point of a circle, here the poloidal section). A configuration (helical
line) is a map 
\[
S^{1}\rightarrow S^{1}
\]
charaterized by an integer number $n$ representing the degree of the map.
This means how many times a circle covers a circle, or, how many times the
internal-space phase variable $\theta $ varies between $0$ and $2\pi $ for a
single turn along the circle representing the base space.
\begin{figure}[htbp]
\hfill\includegraphics[width=2in]{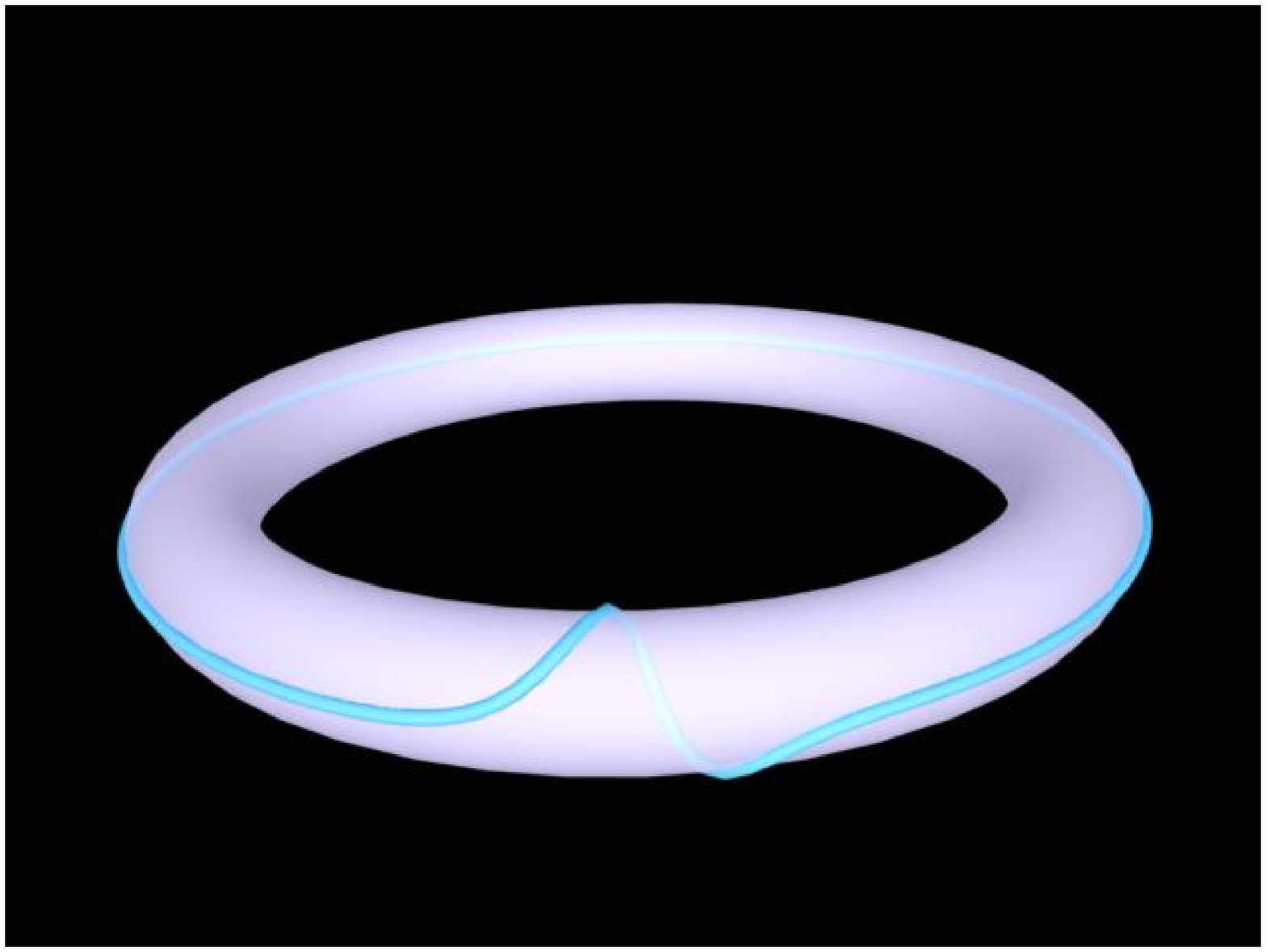}%
\hfill\includegraphics[width=2in]{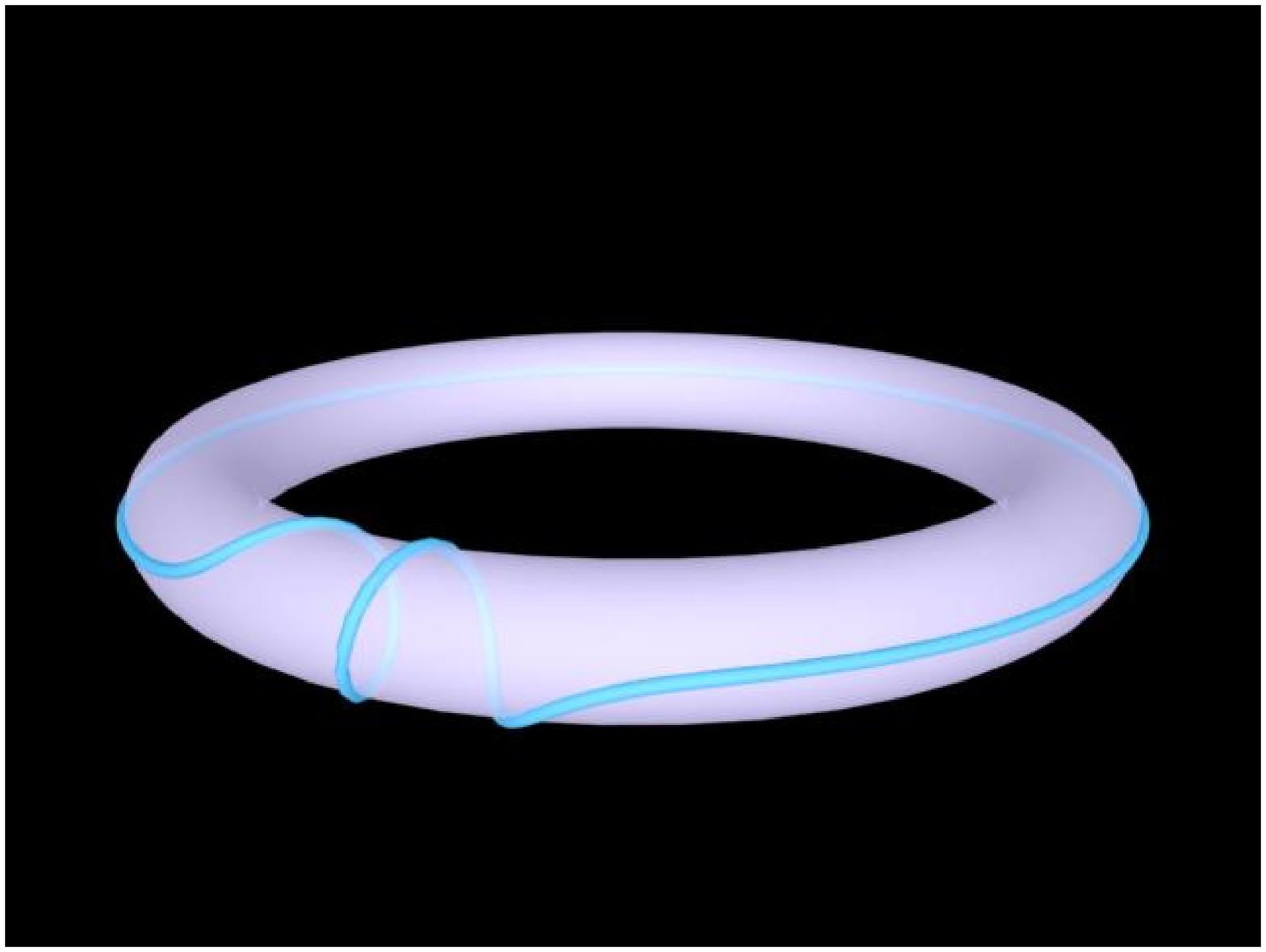}\hspace*{\fill}
\caption{The mapping $S^{1}\rightarrow S^{1} $ with two topological degrees,
$n=1 $ (left) and $n=2 $ (right) } 
\end{figure}

We will look for configurations allowing a transition between a line with
(for example) $n=1$ to another line with $n=2$.

First the model is extended to one where the homotopical deformations are
given by the time dynamics, a much more complex problem, where the simple
lines represent the instantaneous positions of the masses of a chain of
pendul\ae . This is the sineGordon equation.

\subsection{Extending the \emph{sine}-Gordon equation to a larger system}

When we start from an ideal system (like Korteweg-deVries, \emph{sine}%
-Gordon, Nonlinear Schrodinger, Painlev\'{e} transcendents, etc.) and try to
look at it as a reduced form of a larger and more complex system, it
systematically appears the Self-Dual Yang-Mills system.

It has been proved that all the equations mentioned above are reduced forms
of the SDYM equations. The derivation consists of several steps:

\begin{itemize}
\item  definition of the space-time; for the Painlev\'{e} transcendents this
is the complexified four dimensional space $\left( \xi ,\widetilde{\xi }%
,\tau ,\widetilde{\tau }\right) $with metric $ds^{2}=d\xi d\widetilde{\xi }%
-d\tau d\widetilde{\tau }$.

\item  the fibre bundle structure, where the basis is the space-time and the
fibre is a vector space with the group of automorphism $SL\left( 2,\mathbf{C}%
\right) $. The gauge potential is a connection one-form and the self-duality
expresses the equality of the curvature two-form with its Hodge dual.

\item  imposing the invariance to a group of symmetries. The generators of
the algebra of the symmetry group are constructed using the normal form of
the matrix that combines the basis vector fields in the projective
coordinates.

\item  the connection one-form is contracted with these generators and
particular expressions are obtained. Requiring self-duality generates the
nonlinear equations
\end{itemize}

The abstract procedure can be expressed in the language of twistors.

After finding that the idealized form of the initial system can be seen as a
reduced form of the SDYM system, we can look for a sufficiently general
framework for the SDYM : larger dimensionality, larger symmetry group, etc.
The aim is to reach the form of the theory where the solution connecting the
two topological structures is available. In this idea, the extension of the
SDYM system to the Hitchin system will prove to be essential for identifing
the passage between the two $S^{1}\rightarrow S^{1}$ states characterized by
different topological degrees $n$ as the Riemann surface representing
interaction of two strings.

\subsection{String theory realisation of the instanton as a Riemann surface}

Looking for systematic methods to construct instantons connecting
configurations with different topologies, we must start with the cases where
the initial and final states have the simplest nontrivial homology. These
are essentially the classes of states of a system representing the mapping $%
\varphi $ 
\[
S^{1}\stackrel{\varphi }{\rightarrow }S^{1} 
\]
whose first homotopy group is $\mathbf{\pi }_{1}\left( S^{1}\right) =\mathbf{%
Z}$. Such a system is the \emph{sine}-Gordon equation. The states are
classified by the integer $n$ and we ask how to connect, for example, a
state $n$ with a state $n^{\prime }\neq n$. We can find a suggestive
algebraic-geometric construction in the \emph{Matrix theory}.

In general terms the Matrix theory is the non-perturbative description of
the $M$-theory which, in turn, is the $11$-dimensional theory of the strong
coupling limit of type $IIA$ strings. The string theory can be obtained from
the $M$-theory by compactification on a circle and identifying the string
coupling constant with the radius of this circle. By compactification it
results a supersymmetric Yang-Mills theory (SYM) which describes
non-perturbatively the type $IIA$ strings. The duality of these theories
(strings and SYM) is underlined by the relation between their coupling
constants $g_{s}=\left( \sqrt{\alpha ^{\prime }}g_{SYM}\right) ^{-1}$ and
means that examining the strong coupling limit of SYM we get the weakly
interacting, \emph{i.e.} perturbative, limit of the string theory. In Ref. 
\cite{DVV} it is shown that the deep infrared limit (equivalent to strong
coupling in SYM) is a theory describing strings propagating freely. The
supersymmetric Yang-Mills theory offers two models equally interesting for
their topological properties. In the absence of interaction (a phase with
completely broken $U\left( N\right) $ symmetry), the strings are multiply
wound along the compact dimension of the cylindrical base space. When a weak
interaction is allowed we have separation and connection of strings. At this
point a Riemann surface is introduced and this is the construction we want
to examine. However, in order to reach that point we have to say few words
about the \emph{matrix} theory. Everything that follows can be found in the
papers related to the $M$-theory, in particular in Refs. \cite{DVV}, \cite
{Wynter}, \cite{Bonelli} , which we strongly recommend to be read for more
detailed explanation of this framework.

The starting point is the supersymmetric Yang-Mills theory in $10$
dimensions. This is dimensionally reduced to two spatial dimensions yielding
a SYM theory with gauge group $U\left( N\right) $ defined on the $1+1$
dimensional Minkowski space, with the action 
\begin{eqnarray}
S &=&\frac{1}{2\pi }\int d\tau d\sigma Tr\left[ \frac{1}{2}\left( D_{\alpha
}X^{I}\right) ^{2}+i\Theta ^{T}\not{D}\Theta -\frac{g_{s}^{2}}{2}F_{\alpha
\beta }^{2}\right.  \label{actiune1} \\
&&\left. +\frac{1}{2g_{s}^{2}}\left[ X^{I},X^{J}\right] ^{2}+\frac{1}{g_{s}}%
\Theta ^{T}\gamma _{i}\left[ X^{I},\Theta \right] \right]  \nonumber
\end{eqnarray}
All fields are hermitian matrices of order $N\times N$. The indices $\alpha $
and $\beta $ take values $(0,1)$ and $I$ takes values between $1$ and $8$.
The covariant derivative is $D_{\alpha }X^{I}=\partial _{\alpha }X^{I}+i%
\left[ A_{\alpha },X^{I}\right] .$ The operator of covariant derivative is
contracted with the \emph{gamma }matrices in $2D$ , which verify the
relations: $\left\{ \rho _{\alpha },\rho _{\beta }\right\} =-2\eta _{\alpha
\beta }$ where $\eta _{\alpha \beta }$ is the flat Minkowski metric. The $%
\Theta $ fields consists of $8$ matrices $N\times N$ having as elements the $%
2$-spinors $\Theta ^{T}=\left( \theta _{s}^{-},\theta _{c}^{+}\right) $
where the $\pm $ sign correspond to the chirality in $2$ dimensions and $%
\theta _{s}^{-},\theta _{c}^{+}$ are spinors in the representations $\mathbf{%
8}_{s}$ and\ $\mathbf{8}_{c}$ of $SO\left( 8\right) $. The matrices $\gamma
_{i}$ are $16\times 16$ gamma matrices of $SO\left( 8\right) $. The coupling
constant is $g_{s}$.

For small $g_{s}$ the strings are weakly coupled. At the limit $g_{s}=0$
there is no interaction and all matrices commute. In this case the matrices $%
X$'s and the fermionic fields $\Theta $ can be written 
\[
X^{I}=Ux^{I}U^{\dagger } 
\]
\[
\Theta =U\theta U^{\dagger } 
\]
where $x^{I}$ and $\theta $ are diagonal matrices and the matrix $U$ is
unitary.

\subsubsection{Multiply wound strings}

It is possible to find field configurations corresponding to strings
multiply wound around the compact direction $\sigma $. To make them more
explicit we have to take the matrix $U$ of the form 
\[
U\left( \sigma +2\pi \right) =U\left( \sigma \right) g 
\]
which means 
\[
x\left( \sigma +2\pi \right) =gx\left( \sigma \right) g^{\dagger } 
\]
where $g$ is an element of the Weyl group of the group $U\left( N\right) $.
One can see that the variation around the compact coordinate $\sigma $
yields an interchange of the eigenvalues which form cycles of different
lengths. \ Considering a cycle of length $n$, it implies $n$ eigenvalues $%
x_{1}\left( \sigma ,\tau \right) $, $x_{2}\left( \sigma ,\tau \right) $,
..., $x_{n}\left( \sigma ,\tau \right) $ with the cyclying property 
\[
x_{i}\left( \sigma +2\pi ,\tau \right) =x_{i+1}\left( \sigma ,\tau \right) \;%
\text{et}\;x_{n+1}=x_{1} 
\]

In the infrared limit there are sectors corresponding to different ways to
divide the total number of eigenvalues in cyclic groups 
\[
N=\sum_{n}nN_{n} 
\]
where $N_{n}$ is the number of cycles of length (\emph{i.e.} number of
eigenvalues in the cycle) $n$. For these sectors the original non-abelian
symmetry is broken to the discrete symmetric group $S_{N}$, which has two
consequences: (1) it permutes the different cycles of the same length $n$;
and (2) it performs cyclic permutations inside every cycle.

The string interaction appears when two eigenvalues, as functions of $\left(
\sigma ,\tau \right) $, come close and are interchanged. In the point $%
\left( \sigma ,\tau \right) $ where they are touching a group $U\left(
2\right) $ (a subgroup of the original gauge symmetry $U\left( N\right) $
which was completely broken) is restored.

In conclusion the strings do not interact in the infrared limit, where the
gauge symmetry is completely broken down to the maximal torus $U\left(
1\right) ^{r}$ where $r$ is the rank of the gauge group.

\subsubsection{String interaction}

As we have said, leaving the infrared limit and allowing a weak interaction
it appears in the SYM theory that certain non-abelian subgroups are restored
in some region of the spae-time. This corresponds to the fact that, around a
particular point $\left( \sigma ,\tau \right) $ two eigenvalues of $X$
become equal 
\[
x^{I}=x^{J} 
\]
restoring a $U\left( 2\right) $ symmetry out of two $U\left( 1\right) $.
This means that for a nonzero $g_{s}$ it occurs a transition between states
characterized by the transposition of these two eigenvalues. This is an
elementary process of separation or of connection of two strings.

Let us start from the state where there are only free strings. This means
that the fields $X^{I}$ are diagonal having on the diagonal groups of
eigenvalues forming cycles of various lengths. The fields are 
\begin{equation}
A_{\mu }=0  \label{azero}
\end{equation}
\begin{equation}
X\left( \sigma ,\tau \right) =diag\left( x_{1}\left( \sigma ,\tau \right)
,...,x_{n}\left( \sigma ,\tau \right) \right)  \label{xzero}
\end{equation}
with the cycling condition 
\[
x_{i}\left( \sigma +2\pi ,\tau \right) =x_{i+1}\left( \sigma ,\tau \right) 
\]

The matrix $X$ verifies the ``free'' equation $\partial _{\mu }\partial
^{\mu }X=0$.

When we turn around the interaction point $\left( \sigma ,\tau \right) $ the
eigenvalues are interchanged and the fields are gauge-transformed 
\begin{equation}
X=U\left( \sigma ,\tau \right) diag\left( x_{1}\left( \sigma ,\tau \right)
,...,x_{n}\left( \sigma ,\tau \right) \right) U^{\dagger }\left( \sigma
,\tau \right)  \label{xu}
\end{equation}
\begin{equation}
A_{\mu }=igU^{\dagger }\left( \partial _{\mu }U\right) ,\;\;A_{\tau }=0 
\nonumber
\end{equation}
The fields now verify the non-abelian equations 
\[
D_{\mu }D^{\mu }X=0 
\]
The gauge matrix $U$ has the condition 
\[
U\left( \sigma +2\pi ,\tau \right) =U\left( \sigma ,\tau \right) g 
\]
where $g$ is the cyclic shift matrix 
\begin{equation}
g=\left( 
\begin{array}{ccccc}
0 & 1 & 0 & \cdots & 0 \\ 
0 & 0 & 1 & \cdots & 0 \\ 
\vdots &  &  &  & \vdots \\ 
1 & 0 & \cdots & 0 & 0
\end{array}
\right)  \label{gcycle}
\end{equation}
We conclude that the presence of cycles, \emph{i.e.} multiply wound strings
is connected to the existence of a non-nul pure gauge field $A_{\mu }$. On
the basis cylinder we can change to complex coordinates $\left( \sigma ,\tau
\right) \rightarrow \left( z,\overline{z}\right) $ where $z=\exp \left[ 
\frac{1}{2}\left( \tau +i\sigma \right) \right] $. Then the field $X\left(
\sigma ,\tau \right) $ can be seen as a covering of the complex plane. If
there are no cycles, \emph{i.e.} all eigenvalues of $X$ are distinct (cycles
have length $1$) then $X$ consists of a $N$ distinct sheets covering of the
complex $z$ plane. When multiply wound strings exists there are branching
points of different order placed in the origin. The order of the branching
is the length of the cycle and the length of the string. The \textbf{%
interactions} arise when the branching points are different from the origin.

The explicit determination of the matrix $X$ is based on the fact that it is
a Riemann surface realizing a $N$ -sheet covering of the complex plane,
which means that it is the solution of a polynomial equation of degree $N$

\begin{equation}
\sum_{j=0}^{N}a_{j}\left( z\right) X^{J}=0  \label{xpoly}
\end{equation}

The essential onservation is that the equation (\ref{xpoly}) is satisfied by
the matrix 
\begin{equation}
M=\left( 
\begin{array}{ccccc}
-a_{N-1} & -a_{N-2} & \cdots & -a_{1} & -a_{0} \\ 
1 & 0 & \cdots & 0 & 0 \\ 
0 & 1 & \cdots & 0 & 0 \\ 
\vdots &  &  &  & \vdots \\ 
0 & 0 & \cdots & 1 & 0
\end{array}
\right)  \label{matricem}
\end{equation}

This matrix can be diagonalized by a Vandermonde matrix $S$%
\[
M=S\,diag\left( x_{1},x_{2},\ldots ,x_{N}\right) \,S^{-1} 
\]
\begin{equation}
S=\left( 
\begin{array}{cccc}
x_{1}^{N-1} & x_{2}^{N-1} & \cdots & x_{N}^{N-1} \\ 
x_{1}^{N-2} & x_{2}^{N-2} & \cdots & x_{N}^{N-2} \\ 
\vdots &  &  &  \\ 
1 & 1 & \cdots & 1
\end{array}
\right)  \label{matrices}
\end{equation}
where $x_{i}$, $i=1,\cdots ,N$ are the eigenvalues of $M$ and are scalar
holomorphic functions of $z$, .

Turning around a ramnification point the matrix is multiplied by $g$%
\[
S\rightarrow Sg 
\]
an element of the Weyl group of $U\left( N\right) $. The solution for $S$
can be found explicitely when the surface is known (\emph{i.e.} the
coefficients $a_{i}\left( z\right) $) since the eigenvalues $x_{i}\left(
z\right) $ can be determined. Then $X$ and $A$ can be calculated (an
explicit example for the $\mathbf{Z}_{N}$ covering is given in \cite{Wynter}
and \cite{Bonelli}). It is the nontrivial gauge field configuration $A_{\mu
}=igU^{\dagger }\left( \partial _{\mu }U\right) $ which interpolates between
the winding sector in the past (here: $z=0$) and the winding sector in the
infinite future ($z=\infty $).

\bigskip

In simplified terms this is a generalized form of the $\varphi ^{4}$ theory
or of the model of a particle in a two well potential. Both models can be
embedded into this large framework and their equations can be derived by
reduction from the equations obtained from the action $S$. The states
corresponding in the simple models to mapping the circle onto the circle a
fixed $n$ number of times can be regarded as a cross section of the full
field $X\left( \sigma ,\tau \right) $ at the $g_{s}=0$ limit, although the
content of $X$ can be vastly richer.
\begin{figure}[htbp]
\centerline{\epsfxsize=7cm\epsfbox{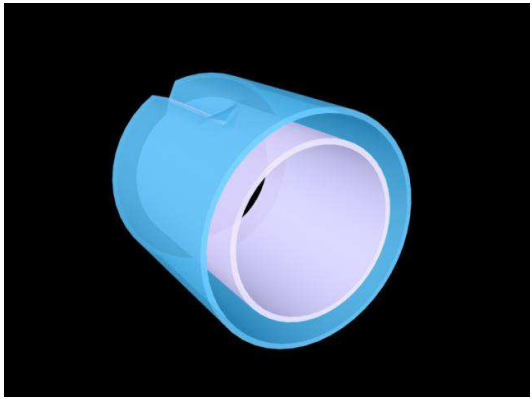}}
\caption{Riemann surface corresponding to the process where two eigenvalues
$X_{1}\left( \sigma, \tau \right) $ and $X_{2}\left( \sigma, \tau \right) $
become equal at a certain $ \tau $. } 
\end{figure}

\begin{figure}[htbp]
\hfill\includegraphics[width=2in]{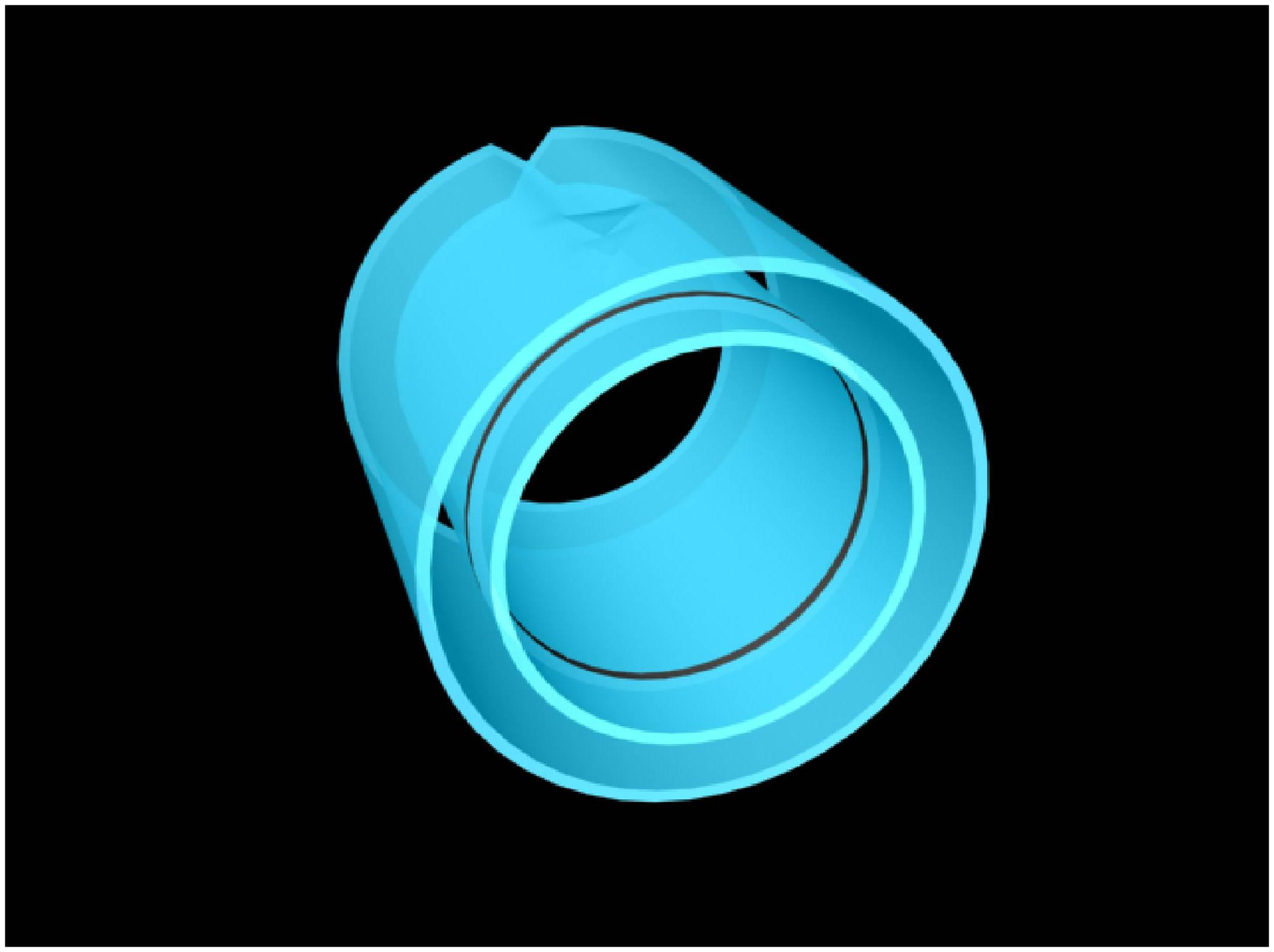}%
\hfill\includegraphics[width=2in]{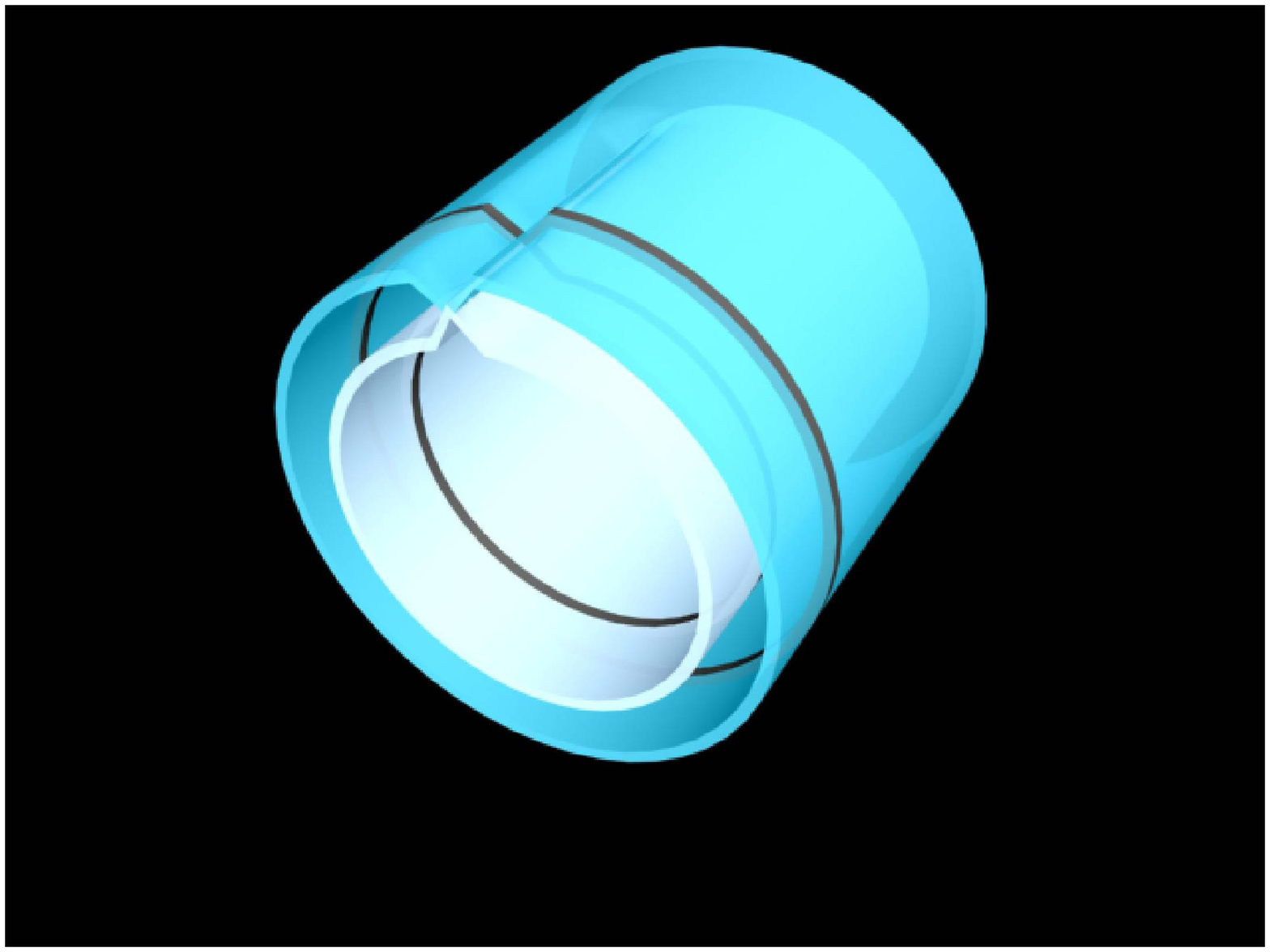}\hspace*{\fill}
\caption{The mapping $\sigma \rightarrow X\left( \sigma ,\tau \right) $ is 
a representation of the topological map $S^{1} \rightarrow S^{1} $. In the
left figure, for times $\tau $ before interaction, the topological degree
is $n=1 $; in the right figure (after interaction) it is $n=2$.} 
\end{figure}

It has been proved above that to the \emph{solution} $X$ of the equations of
motion with weak interaction corresponds a Riemann surface. This allows to
identify the insanton with this surface and to identify the two topological
configurations of different $n$ 's with multiply wound strings in the
incomming and respectively in the outgoing state of the string interaction.

\subsection{Transitions between regimes}

The example of the kink connecting in imaginary time the equilibrium
positions of a classical particle in the two-well potential can be
generalized. The \emph{minima} of the two-well potential is replaced, in
general cases, by \emph{states with stationary properties} corresponding,
for the idealized system, to families with definite topological content. To
the instanton connecting the two particle positions it corresponds in
general a solution connecting these distinct families of states.

Consider the sine-Gordon equation 
\[
\frac{\partial ^{2}u}{\partial t^{2}}-\frac{\partial ^{2}u}{\partial x^{2}}%
+\sin u=0 
\]
In Ref.\cite{Eleonskii} it is found the following solution 
\[
u(x,t)=4\arctan \frac{a\,sn\left( \tau \right) +\exp \left( \Lambda x\right) 
}{1+a\,sn\left( \tau \right) \exp \left( \Lambda x\right) } 
\]
where $a^{2}=k\leq 1$, $\Lambda =\left( 1-k\right) /\left( 1+k\right) \leq 1$%
, $\tau =t/\left( 1+k\right) $. This solution interpolates between two
particular solutions of the sine-Gordon equation (being in this sense
equivalent to an instanton connecting SWSP 's) 
\[
\stackunder{x\rightarrow -\infty }{\lim }u\left( x,t\right) =u_{\omega } 
\]
\[
\stackunder{x\rightarrow \infty }{\lim }u\left( x,t\right) =u_{\omega }+2\pi 
\]
where $u_{\omega }=4\arctan \left( \exp x\right) $. We can see this example
as a particular form of the general structure discussed above.

\subsection{Other possible examples}

\textbf{Solitons as homoclinic curves and intermittency of the Nonlinear
schrodinger Equation}. The intermittency in the case of weakly dissipative,
driven, Nonlinear Schrodinger Equation has been examined numerically \cite
{Ercolani}. It has been found that there are jumps between two kinds of
solitonic solutions on a periodic domain. This has been explained by the
analogy between the soliton and the homoclinic curve of the simple system
like the pendulum, separating two distinct types of behaviour: finite
oscillations and free rotation.

Transitions have been identified between states of complex dynamics, for
systems like sine-Gordon. In some cases, fast changes of the systems between
states of different symmetry patterns have been observed in experiments with
burning gas in porous media, oscillations in Belousov-Zhabotinsky reactions,
etc. Under the same perspective should be examined the models of the
free-force type, in particular the ABC flow. They are known to exhibit
intermittency and also are known to have a self-dual structure.

\section{Discussion}

There are many possible extensions and developments arising from this idea,
some of them being interesting challanges. For example, if this idea is
proved correct and a systematic technical procedure will be available, one
of the most important application will be the description of the
reconnection of vortices in fluids and plasmas (and similar, of the magnetic
structures in weakly resistive plasmas). The approach proposed here
naturally explains why the mere presence of the resistivity is required in
such systems and why its magnitude is less important: most of the time the
fluid performs homotopic deformations and from time to time rapid
reconnection events changes the topological degree. The amount of energy
implied in this event is not significant while the simple presence of the
resistivity is required if we want reconnections to be possible.

A challanging problem which is closely related to our model is the
generalization of the Painlev\'{e} criterion. It is clear that the
singularities of an instanton defined in a much larger theory than the
original system cannot be simply reduced to singularities in the complex
plane and can possibly appear as singularities of Riemann surfaces or as
vanishing cycles. A precise connection between the intagrability of a model
and the singularity structure of the instanton would represent a
generalization of the Painlev\'{e} theory.


\begin{thebibliography}{99}
\bibitem{FrischMorf}  U. Frisch and R. Morf, Phys. Rev A \textbf{23},2673
(1981).

\bibitem{Bassetti}  B. Bassetti, P. Butera, M. Raciti and M. Sparpaglione,
Phys. Rev. A \textbf{30}, 1033 (1984).

\bibitem{Fucito}  F. Fucito, F. Marchesoni,E. Marinari, G. Parisi, L.
Peliti, S. Ruffo and A. Vulpiani, J. Physique \textbf{43}, 707 (1982).

\bibitem{Rajaraman}  Rajaraman, \emph{Solitons and instantons}, North Holland, 1982.

\bibitem{FlorinMadi}  F.Spineanu, M.Vlad, ''Statistics of intermittency for
dissipative systems '', STATPHYS 20, Paris, July 1998 ;

\bibitem{FlorinMadi1}  F. Spineanu, M. Vlad, ''Aspects dynamiques de
l'intermittence dans les syst\`{e}ms dissipatifs'', ''Rencontres de
non-lin\'{e}aire'', 15-16 March 2001, Paris, France, in Rencontre du Non
Lin\'{e}aire, Editors Y. Pomeau and R. Ribotta, (2001) Non Lin\'{e}aire
Publications, Orsey, Paris (ISBN 2-9516773-0-8), 237-242.

\bibitem{Ward}  R. S. Ward, Phys.Lett.\textbf{61A}, 81 (1977).

\bibitem{Mason1}  L. J. Mason and G.A.J. Sparling, Phys.Lett.\textbf{A137},
29 (1989).

\bibitem{Mason2}  L. J. Mason and N. M. J. Woodhouse, Nonlinearity \textbf{6}%
, 569 (1993).

\bibitem{Ercolani}  N. Ercolani, M.G. Forest and D.W. McLaughlin, \ Physica 
\textbf{D43}, 349 (1990).

\bibitem{DVV}  R. Dijkgraaf, E. Verlinde and H. Verlinde, \emph{Matrix
string theory}, hep-th/9703030.

\bibitem{Wynter}  Th. Wynter, \emph{Gauge fields and interactions in matrix
string theory}, hep-th/9709029.

\bibitem{Bonelli}  G. Bonelli, L. Bonora, F. Nesti and A. Tomasiello, \emph{%
Matrix string theory and its moduli spaces}, hep-th/9901093.

\bibitem{Eleonskii}  V. M. Eleonskii, N. E. Kulagin and N. S. Novozhilova,
Sov. Phys. JETP \textbf{62}, 1255 (1985).
\end{thebibliography}
\end{document}